# Visualising Emotional Landmarks in Cities

Salvatore Iaconesi, Oriana Persico
ISIA Design Florence
{salvatore.iaconesi@artisopensource.net, oriana.persico@gmail.com}

**Abstract**

*Different people and cultures associate different emotional states to different parts and spaces of cities.*

*These vary according to individuals, their cultures and also to the time of day, day of week, season, special occasions and more.*

*Recurring patterns may occur in correspondence of the places in which people work, study, entertain themselves, consume, relate, wait or just take a break.*

*What can we learn from these patterns?*

*Trying to find possible answers to this question passes through the possibility to visualize and represent the configurations of emotional expressions in urban spaces, across time, geography, theme, cultures and other dimensions.*

*We have developed ways in which it is possible to harvest people's geo-located (or geo-locatable) emotional expressions from major social networks and to visualize them according to a variety of different modalities.*

*In this paper we will present a series of these types of visualizations, and the ways in which they can be used to gain better understandings of these emotional patterns as they arise, from points of view which derive from anthropology, urbanism, sociology, politics and also arts and poetics.*

*The paper will focus on the ways in which the data is harvested from different social networks, then categorized and annotated with meta-data describing the emotional states, the languages in which people express themselves, the geographic locations, the themes expressed.*

*A methodology for representing this information across a variety of domains (time, space, emotion, theme) will then be presented in detail.*

*A reflection on possible usage cases for anthropology, urbanism, policy-making, arts and design will end the contribution, as well as the description of series of open issues and the indication of possible next-steps for research.*

*Keywords*--- **emotional analysis, social networks, natural language processing, public space, anthropology, urban planning, architecture**.

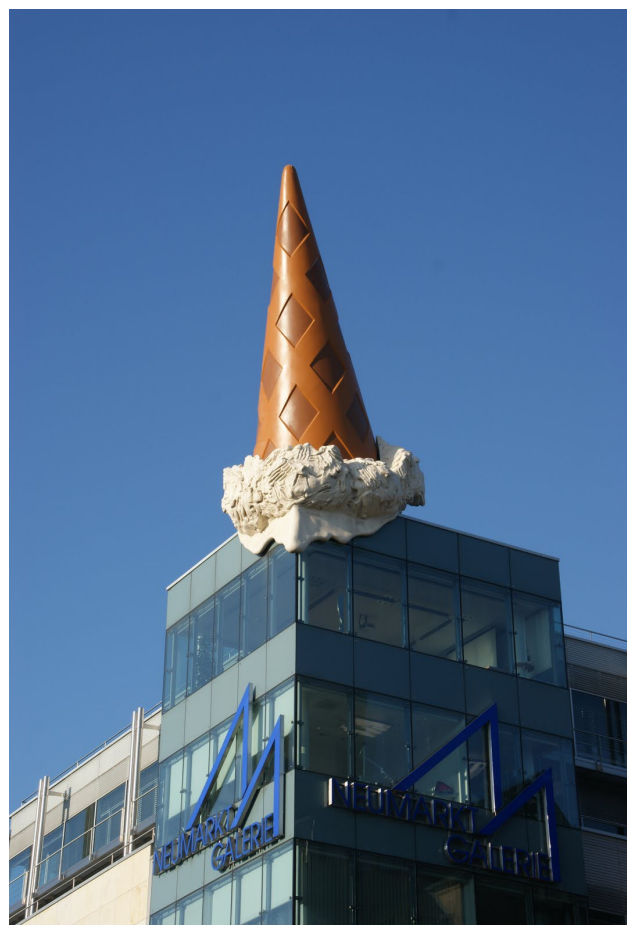

*Figure 1: C. Oldenburg, Dropped Cone, Neumarkt Galerie, Cologne, Germany.*

## 1. Introduction

Neumarkt Square, Cologne, Germany. A giant ice cream come seems to have been dropped atop a steel and glass building at the corner of the plaza (Figure 1).

In a clear, sunny day, while many bicycles crowds swarm about, multiple people are taking pictures of it and of themselves. Many have evidently expecting someone. We can almost imagine their text messages from a short while ago: "Meet me at the giant ice cream cone." For the people of the city, it has become an usual, pleasant view: they notice it with the corner of their eye, and while it has become a common view, it never fails to bring out a smile, or the curiosity of how they brought it up there, or visions of the gigantic child with a sad face that has spilled his cone onto the tiny building that was standing at teh height of his ankles. For the people who arrive for the first time, it is the perfect moment to shoot out a sincere laugh. "Look!" "Did you see it?" and even "That's so stupid!" are some of the most common comments pronounced out loud which you can hear if you are an eavesdropping urban ethnographer.

This artwork by Claes Oldenburg is one of many extraordinary examples of what can be described as emotional landmarks: places in which specific emotional expressions are systematically produced by multiple city dwellers of different backgrounds and cultures.

Emotions are a very important part of our experience of the world, and express the ways in which people perceive a sense of well-being (or the absence of it) in their environment.

How can we observe and visualize human emotions in cities to gather better understandings about their joys, fears, excitements, doubts, wishes and, more in general, expectations and desires?

## 1.1. The Sculptor and the Architect

Cities are both concrete and imaginary.

One side the city is "characterized by rigidity, paralysis — the halting of motion in which the city expresses its architectural and urban riches" [1]. On the other side it is "marked by the threat of imminent explosion due to a precarious creativity that seeks spaces and territories for its expression, thereby producing images and edifices, routes and figurations that reveal the impassioned results of a life lived in art and architecture" [1].

The ephemeral city betrays a latent desire for a dream with an urgent drive to reveal and disseminate itself through reality. What is called "public art" is a residue of the creative exuberance that imbued both the theory and planning of the "ideal city". It is the government's compensation for Utopia that, unable to penetrate reality, was abandoned in the squares and silent territories of the city, to maintain the presence/absence of an alternative dream and desire. [1]

During the 1960s, Claes Oldenburg took up the challenge to restore an ideal urban impact that could compete with the existing cityscape, transforming sculpture in geographic invention, transferring urbanism from one language, architecture, to another, art. Oldenburg multiplies the element of surprise which distorts the laws of construction by imposing a phantasmagorical scene that explores a language of figural excess.

## 1.2. The Ubiquitous Sculptor of the City

With mobile devices, such as smartphones and tablets, we have become able to produce a continuously evolving *infoscape* – the layer of information which forms our experience of the world, together with the physical landscape. Through social networks and other services and tools, we have become able to express our emotions, points of view, desires, expectations and wishes in ways that are able to influence others' actions, decisions and emotions.

Through this production we have become a peculiar type of sculptors, transforming our ephemeral, transient, temporary perceptions and expressions into tangible artifacts, which are disseminated in the environment. Images and texts are constantly published onto public spaces, shopping centers, offices, schools and streets, forming a new layer of reality [2, 3 and 4]. These artifacts contribute to the formation of our everyday experience of cities.

## 2. Emotions, previous work

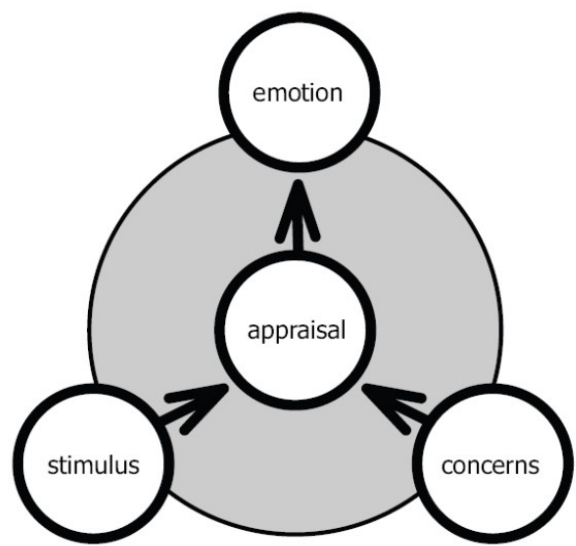

*Figure 2: P. Desmet, Basic model for emotions.*

Desmet [5] has described a basic model of emotions (Figure 2) on top of the work of Rosemann [6], Ortony [7] and Lazarus [8].

In this model, the concept of Appraisal is defined as a non-intellectual, automatic evaluation of the significance of a stimulus for one's personal well-being. The act of emotional appraisal leverages the presence of two different inputs: Stimulus and Concerns. According to Frijda [9] a Stimulus is any perceived change which has the potential to elicit an emotion. Every emotion hides a concern, that is, a more or less stable preference for certain states of the world. Thus [9] concerns can be regarded as points of reference in the appraisal process. It is interesting to note Concerns' relation to Maslow's [10] hierarchy of needs: Concerns can be placed at the different levels of the hierarchy to express their referring domains, their importance and priority.

Multiple methodologies for emotional appraisal of architectural environments exist.

Some [11, 12] focus on complexity as a parameter for evaluation, referring to the number and the varieties of different units present in a setting/picture. Kaplan [13] proposed a model on environmental preference in which predictors like complexity play an important role. According to Berlyne [12] different levels of complexity can be associated to different levels of preference.

Familiarity and prototypicality can influence the emotional reaction to architectural spaces [14].

According to the circumplex aesthetic model proposed by Russell [15] the affective appraisal of the environment can be summarized by two dimensions: pleasantness and arousal. In view of another aesthetic model called Discrepancy Model [16], the level of likeableness attributed to an external stimulus depends on how far is the appraised environment from the prototypical exemplar that an individual has in mind. If the incoming stimulus is too similar or too different from the prototype, it is very possible that the evaluation will be negative; on the contrary at a moderate level of discrepancy there are good chances that the stimulus will be appreciated.

Some methodologies are focused on the possibility to understand how emotions are expressed in the daily lives of citizens.

Christian Nold's fundamental work on Biomapping [18] and Emotional Cartography [19] saw multiple people in 25 cities of the world wear Galvanic Skin Response devices to record their emotional arousal in conjunction with their geographical location. In this way, a map is created which visualizes points of high and low arousal.

Another example, the Fuehlometer ('feel-o-meter') [19], was produced by german artists Wilhelmer, Von Bismarck, and Maus in the form of a public face, an interactive art installation that reflects the mood of the city via a large smiley face sculpture. It was installed atop a lighthouse in Lindau, Germany. A digital camera along the lake captured the faces of passersby, which were then analyzed by a computer program and classified as either happy, sad, or indifferent. The cumulative results determine the expression of the sculpture, whose mouth and eyes shift accordingly via a system of automated motors.

Using a different approach, the City of Vilnius [20] used a social tool to gauge the average residents' level of happiness, using an online form and barcodes in the city.

Another example comes from Consciousness of Streams [21], in which art installations in multiple cities were used to contribute users' emotional state and geographical location, creating a "real-time interconnected emotional map of the planet" [22].

The Mappiness project [23], part of a research project at the London School of Economics, created a mobile app which contacted users once a day to ask how the were feeling. The resulting data would be used to create information visualizations.

The Aleph of Emotions [24], an experimental art project by Mithru Vigneshwara, used a camera-like interface allowing users to point along a particular direction, focus to a place along that direction, and click to view a representation of emotions in that place. Emotions were collected by searching for keywords on social networks.

## 3. Design

From the case studies we were able to form an idea of what the previous work in the domain had been able to achieve and the major critiques relative to the type of result we set forth to obtain.

Among the issues we have identified are the idea of emergence and of the unconscious.

Apart from the feel-o-meter, which focuses on facial recognition from people in the streets of the city, all of the other works try to establish a relationship with people, to be able to explicitly ask for their emotional state and geographical position. While this is an interesting approach, it evades our most pressing need: fining ways in which to measure emotions in the city (and, thus, the mental maps according to which people perceive urban space) in emergent form. What we were trying to do was to be able to sense the unconscious elements of people's perception, to infer how people feel in the city by gaining better understandings of their daily activities and expressions, not through the synthesis of a survey, however advanced or technological it might be.

A project which went in this direction was the Aleph of Emotions. One drawback in its approach was its extreme simplicity: keyword based emotional evaluation is not reliable and leaves out most of the irony, creativity and other modalities in which emotions can be expressed. To express "fear" one doesn't always mention the word "fear".

For this we set up a more complex system for information harvesting and processing. We combined the system used for harvesting content from geographical areas described in [25] with the Natural Language Analysis technique described in [26, 27 and 28].

We approached the possibility to recognize emotions by identifying in text the co-occurence of words or symbols (for example smileys) that have explicit affective meaning. As suggested in by Ortony et al. [26] we must separate the ways in which we handle words that directly refer to emotional states (e.g.: fear, joy) from the ones which only indirectly reference them, based on the context (e.g.: "killer" can refer to an assassin or to a "killer application"): each has different ways and metrics for evaluation.

For this, we have used the classification found in WordNet Affect[29], which is an extension of the WordNet database.[30]

The approach we used was based on the implementation of a variation of the Latent Semantic Analysis (LSA). LSA yields a vector space model that allows for a homogeneous representation (and hence comparison) of words, word sets, sentences and texts.

According to Berry [27], each document can be represented in the LSA space by summing up the normalized LSA vectors of all the terms contained in it.

Thus a *synset* in WordNet (and even all the words labeled with a particular emotion) can also be represented in this way.

In this space an emotion can be represented at least in three ways: (i) the vector of the specific word denoting the emotion (e.g. "anger), (ii) the vector representing the synset of the emotion (e.g. {anger, choler, ire}), and (iii) the vector of all the words in the synsets labeled with the emotion.

This procedure is well-documented and used, for example in the way shown in [28], which we adopted for the details of the technique.

We proceeded to harvest the data, and to use an annotation system to mark it with the emotional expressions which we progressively found using this methodology.

### 3.1. Visualizations

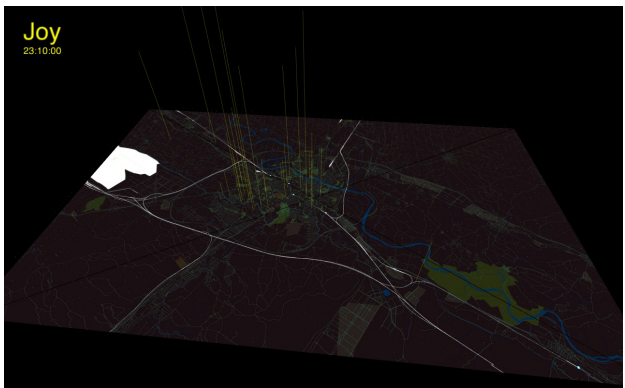

*Figure 3: S. Iaconesi, O. Persico. 3D emotional map in Rome.*

After several preliminary studies in which we used 3D histograms and surfaces to understand the quality, and distribution of the data, we proposed three types of visualizations: a 3D map, a relational visualization and an emotional compass.

The 3D map (Figure 3) highlighted using different colors the different emotions which appeared on social networks. Filters were used to turn on and off the various emotions and languages (we harvested content in 29 languages using the available Wordnet translations). A timeline was used to navigate through time, to observe how emotional expressions moved and appeared/disappeared in the city.

This was a highly legible visualization, as it allowed us to clearly identify the emergence of Emotional Landmarks: geographical locations in which multiple people from different cultures and backgrounds consistently express specific emotions (eventually more than one) at specific times of day/week/season/year. These are places that constitute particularly strong stimuli for people's emotional appraisal. They can (and usually are) temporary (but, eventually, recurrent), and depend on the time of the day (they might be different in office hours or on week-ends, for example) and of the cultural background (i.e.: in the city of Florence different locations are the sites for very different emotional expression from the tourists coming from may different parts of the world). They might be positive or negative (for example in the Schuster Park in Rome, which is used by many people as a shortcut from the bus station to the houses which are on the other side of it, which at night is a place of anxiety or even fear for multiple people, because of its darkness, right in downtown Rome).

This visualization brought up the idea of being able to "ask questions to the city": a modality for interrogating the city not about the typical questions which are often asked ("where is the bus stop?", "where is the museum?", and more), but entirely new types of questions, whose answers can only come from the expressions of high number of people ("where are people afraid?", "where do people feel joy on monday morning?", "where are people anxious, or bored on Friday evening?", and more along similar schemes).

For this reason we set forth in designing the relational visualization and, then, the emotional compass.

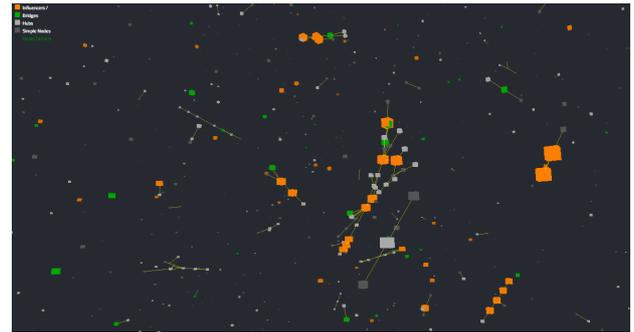

*Figure 4: S. Iaconesi, O. Persico. Relational view of the city.*

The relational visualization (Figure 4) was created to gain better understandings of how people's emotions were connected according to time, place and theme. Conversation threads were aggregated for this, describing discussion communities (non-linear graphs message-reply chains). These were then displayed in 3D so that the people's roles in conversations could be highlighted. This was specifically useful to identify the origins of specific emotions, in their being discussed along the evolving conversations. Many people, for example, showed up as not actually feeling the emotion manifested in a certain space, but simply commenting it from another location, or acknowledging it because of their relationship with another social network user (for example through a "like" on Facebook, or a re-tweet on Twitter).

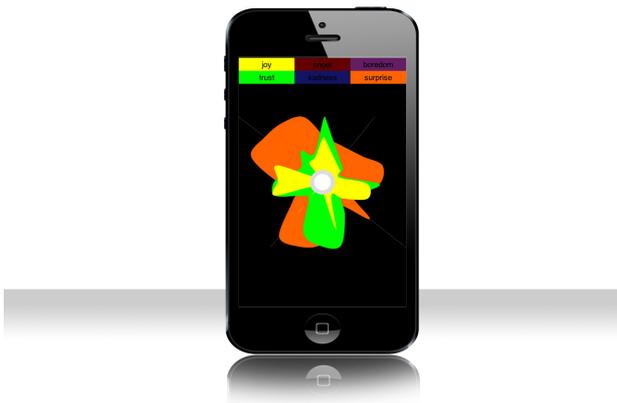

*Figure 5: S. Iaconesi, O. Persico. The Emotional Compass.*

Finally, the Emotional Compass visualization was designed, under the form of an interactive mobile interface for an innovative form of urban navigation (Figure 5). The interface could be used as a standard compass, with the exception that it did not show the direction of cardinal signs, but, rather, the directions in which emotions were expressed on social networks. If the user selected, for example, the emotion of "joy", a color coded shape would appear in radial form, which was thicker in the directions in which more expressions of "joy" were published on social networks, allowing the user to go in that direction to find these areas of the city. Threshold levels were defined to allow users to identify the presence of an emotional landmark. If a certain highlighted emotion was manifested in the current location beyond the threshold level the color of the screen would change to reflect that an Emotional Landmark had been reached. Filters and timeframes could be specified on the interfaces, to allow users to ask more complex questions, in the modalities explored in the previous sections.

## Conclusions

This research has proven to be truly insightful both for providing a methodology using which it is possible to observe human emotions as they unfold in urban spaces, and for some of the results and outcomes which have been collected.

The idea of being able to define Emotional Landmark has turned out to be interesting and deserving further investigation. Their variability across times, cultures and contexts brings up interesting possibilities to understand more about our urban environments, and about the ways in which multiple cultures traverse them for work, entertainment, relations and consumption. Emotional Landmarks exist and are identifiable, giving rise to entirely new opportunities for anthropology, ethnography, sociology as well as for the development of novel services, businesses and possibilities of great utility for administrations and organizations.

Possibly most important of all, the possibility to "ask questions to the city" has presented itself with great emphasis. The opportunity to pose questions to which large numbers of people are able to contribute their answers from emergent, subjective points of view through the practice of their daily lives and activities, and to analyze them with the help of expressive, readable information visualizations is a fertile ground to gain precious insights about our well-being in cities.